\documentclass[fleqn,12pt,twoside]{article}
\usepackage{espcrc1}
\usepackage{epsf}

\usepackage{graphicx}
\usepackage[figuresright]{rotating}

\newcommand{\AmS}{{\protect\the\textfont2
  A\kern-.1667em\lower.5ex\hbox{M}\kern-.125emS}}

\newcommand{\be}{\begin{equation}}
\newcommand{\ee}{\end{equation}}
\newcommand{\beq}{\begin{eqnarray}}
\newcommand{\eeq}{\end{eqnarray}}

\title{ Hadron wave functions and the issue of nucleon
deformation~\thanks{Talk presented by C.~Alexandrou.}}
 
\author{C. Alexandrou\address[Cyprus]{Department of Physics, University of Cyprus,
CY-1678 Nicosia, Cyprus},
Ph.\ de Forcrand\address{ETH-Z\"urich, CH-8093 Z\"urich and CERN, 
CH-1211 Geneva 23, Switzerland}
and A. Tsapalis\addressmark[Cyprus]}
       
\begin{document}

\maketitle

\begin{abstract}
 Using  gauge invariant hadronic two- and three- density correlators
we extract  information on
the spatial distributions of
quarks  in hadrons, and on hadron shape and multipole moments within quenched
lattice QCD.
Combined with the calculation of N to $\Delta$ transition amplitudes
the issue of nucleon deformation can be addressed.
\end{abstract}

 
\section{Introduction}
Correlation functions calculated in lattice QCD can be connected to
basic hadronic features. Two- and three- density correlators for mesons
and baryons reduce in the nonrelativistic limit to the square
of the wave function and therefore provide
detailed information
on hadron structure 
 such as quark spatial distributions,
hadronic shapes and charge radii.
In addition lattice results can be used to test predictions in various models.
Direct connection to the issue of nucleon deformation,
currently under experimental
investigation~\cite{Bates,Glass}, is made by calculating
the {$\gamma^* N$ to $\Delta$} transition form factors in quenched
and unquenched lattice QCD. The aim is to obtain an accurate determination
of the  ratio 
of the electric quadrupole to the magnetic dipole 
amplitudes, 
$R_{EM}(q^2)$, as a function of the momentum
transfer $q$.
A non-zero {$R_{EM}$} arises due to deformation in the
 nucleon and/or $\Delta$ and it is attributed to different mechanisms
in the various models: 
in quark models the deformation arises  due to the colour-magnetic tensor
force whereas
in 'cloudy' baryon models  due to the meson exchange currents.

Let us first consider the
equal-time correlators~\cite{AFT}, 

\vspace{-0.3cm}

\be
 C_\Gamma^H({\bf r},t) = \int\> d^3r'\>
\langle H|\hat{\rho}_\Gamma^u({\bf r'},t)\hat{\rho}_\Gamma^{d}({\bf r}'+{\bf r},t)|H\rangle \quad,
\ee

\begin{figure}[h]
\vspace*{-0.7cm}
\begin{minipage}[b]{6.5cm}
\epsfxsize=6truecm
\epsfysize=5truecm
\mbox{\epsfbox{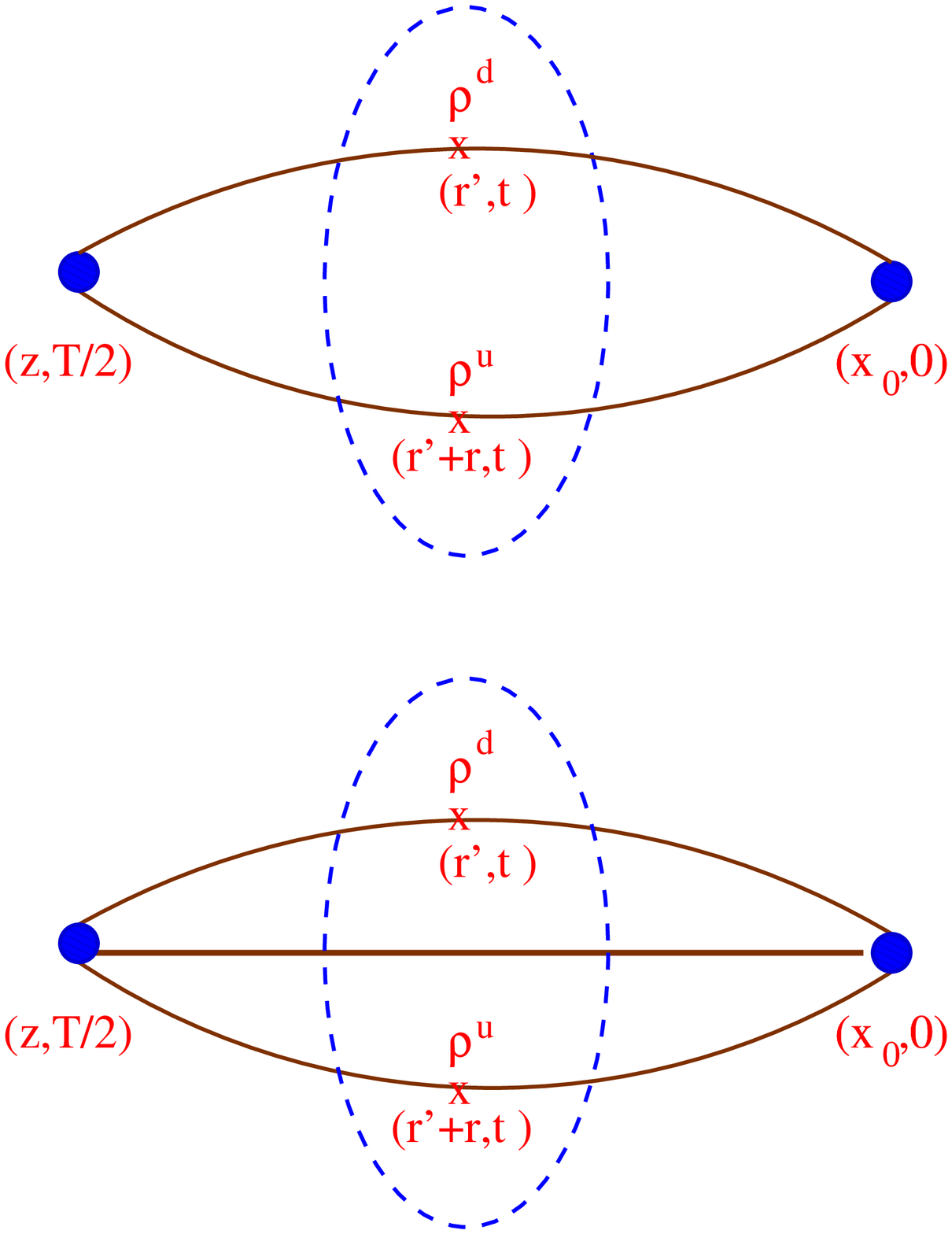}}
\vspace*{-1cm}
\caption{Equal-time correlator for a meson (upper)
and a baryon (lower).}
\label{fig:meson}
\end{minipage} 
\hspace{\fill}
\begin{minipage}[b]{8.5cm}
\epsfxsize=8truecm
\epsfysize=5truecm
\mbox{\epsfbox{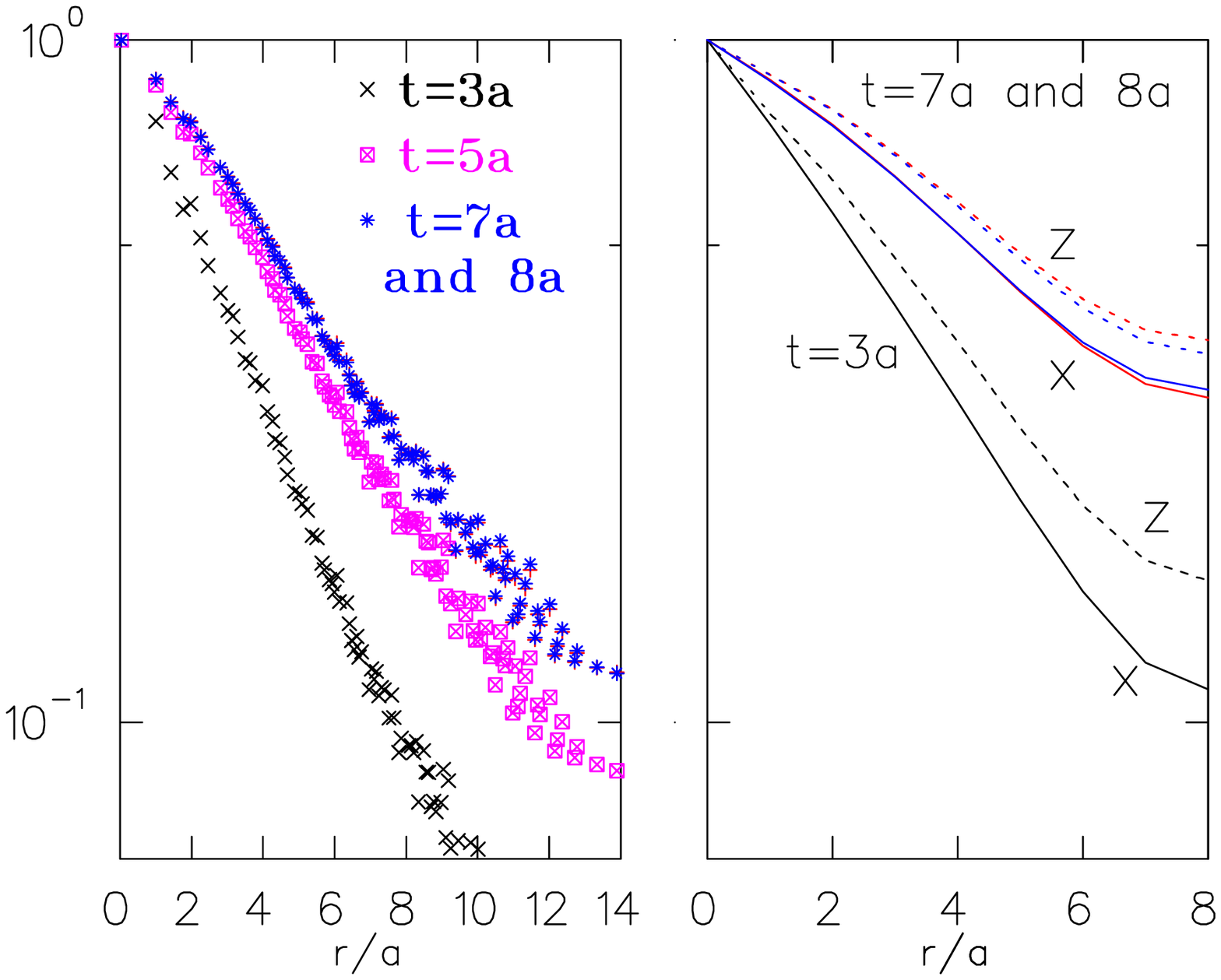}}
\vspace*{-1cm}
\caption{Time evolution to  ground state for the rho  meson at $\kappa=0.153$.}
\label{fig:rho it}
\end{minipage}
\vspace{-0.8cm}
\end{figure}

\vspace{-0.3cm}

\noindent
with $\hat{\rho}^u_\Gamma({\bf r},t)$ given by the normal order product
$:\bar{u}({\bf r},t)\Gamma u({\bf r},t):$.
For $\Gamma =\gamma_0$ and 
 $\Gamma ={\bf 1}$ we obtain, in the nonrelativistic limit,
the charge and matter density 
respectively. For $\Gamma=\gamma_5$ we obtain
the pseudoscalar density which we will compare to bag model predictions.
The  matrix elements of
two-density correlators are shown schematically in Fig.~\ref{fig:meson}.
The advantage of using
density correlators is that they are gauge-invariant unlike Bethe-Salpeter amplitudes.
For baryon wave functions three-density correlators are needed. Here we
will only show results for the one-particle density obtained
after integrating the wave function over one relative distance.
The diagram involved is shown in the lower part 
of Fig.~\ref{fig:meson}.

\noindent

Since the spectroscopic quadrupole moment of the nucleon vanishes
in order to determine any possible deformation we evaluate the 
N to $\Delta$ electromagnetic transition form factors.
We calculate  the three-point correlation
functions~\cite{all} $\langle G^{\Delta j^\mu N}_{\sigma} 
(t_2, t ; {\bf p}^{\;\prime}, {\bf p}; \Gamma) \rangle $
and  \\
$\langle G^{N j^\mu \Delta}_{\sigma} (t_2, t ; {\bf p}^{\;\prime}, {\bf p}; \Gamma) \rangle $ and
 consider the  following ratio~\cite{Leinweber}

\vspace*{-0.5cm}

\beq
R_\sigma (t_2, t; {\bf p}^{\; \prime}, {\bf p}\; ; \Gamma ; \mu)& =&
\large{\left[ \frac{
\langle G^{\Delta j^\mu N}_{\sigma} (t_2, t ; {\bf p}^{\;\prime}, {\bf p};
\Gamma ) \rangle \;
\langle G^{N j^\mu \Delta}_{\sigma} (t_2, t ; -{\bf p}, -{\bf p}^{\;\prime};
\Gamma^\dagger ) \rangle }
{
\langle -g_{ij} G^{\Delta \Delta}_{ij}(t_2,{\bf p}^{\; \prime};
\Gamma_4) \rangle \;
\langle G^{NN} (t_2, -{\bf p} ; \Gamma_4) \rangle } \right]^{1/2}
} \nonumber \\
 &\>&\hspace{-4cm}\stackrel{\Longrightarrow}{t_2 -t \gg 1,t \gg 1}\hspace{1cm} \biggl( \frac{E_N({\bf p})+M_N}{2E_N({\bf p})} \biggr)^{1/2} \biggl( \frac{E'_\Delta({\bf p}')+M_N}{2E'_\Delta({\bf p}')} \biggr)^{1/2} \bar{R}_{\sigma}({\bf p}^{\; \prime}, {\bf p}\; ; \Gamma ; \mu)
\label{R-ratio}
\eeq
from which the transition form factors can be extracted.
The index $\sigma$ is the Lorentz index for 
a spin- $\frac{3}{2}$ field,
$j^\mu (x)$, is the lattice conserved electromagnetic current inserted at
time t,
$E_N({\bf p})$ and $ E'_{\Delta}({\bf p'})$ are the energies of the nucleon and
the $\Delta$ respectively and $t_2$, the time location of the sink,  is varied
to identify the ground state.
We use  projection matrices $\Gamma_j= 1/2{\sigma_j \> \>0 \choose 0 \>\> 0 }  $
and $\Gamma_0=1/2{I \> \>0 \choose 0 \>\> 0 }$
as  in ref.~\cite{Leinweber}.

\vspace*{-0.4cm}

\section{Results}


We have analysed 220 quenched configurations 
at  $\beta=6.0$ for a lattice of size $16^3\times 32$
at $\kappa=~0.15,~0.153,~0.154,~0.155$ which give ratio of pion to rho
mass of $0.88, \> 0.84,\>0.78,\>0.70$ respectively. 
Using the standard definition of the 
 naive quark mass,
$2m_q a= (1/\kappa-1/\kappa_c)$, where $\kappa_c$    
 is the value of $\kappa$ at which
 the pion becomes massless, we find $m_q \sim 300, 170 ,130$ and $85$~Mev
respectively.
We   used the value of the string tension
to set the physical scale obtaining for the inverse lattice
spacing $a^{-1}\sim 1.94$~GeV. 
The density insertions are taken at $t=T/4=8a$.
 To check that this time
interval is long enough to extract the physical hadronic ground
state we computed  the density-density
correlators for varying values of the insertion time $t$. 
Fig.~\ref{fig:rho it} shows the results for the rho correlators which
converge  when $t > 6a$.
Moreover, we note a clear deformation (ratio 
$C_{\gamma_0}^\rho(x,0,0) / C_{\gamma_0}^\rho(0,0,z) \neq 1$)
which remains approximately 
the same as early as $t=3a$ even 
 though the $z$- and $x$-profiles change appreciably
showing that the  deformation is a robust, 
physical property of the rho meson in its ground state as well as its
low-lying excited states.

\begin{figure}[h]
\begin{minipage}[b]{7.8cm}
\vspace*{-1.2cm}
\epsfxsize=8truecm
\epsfysize=7truecm
\mbox{\epsfbox{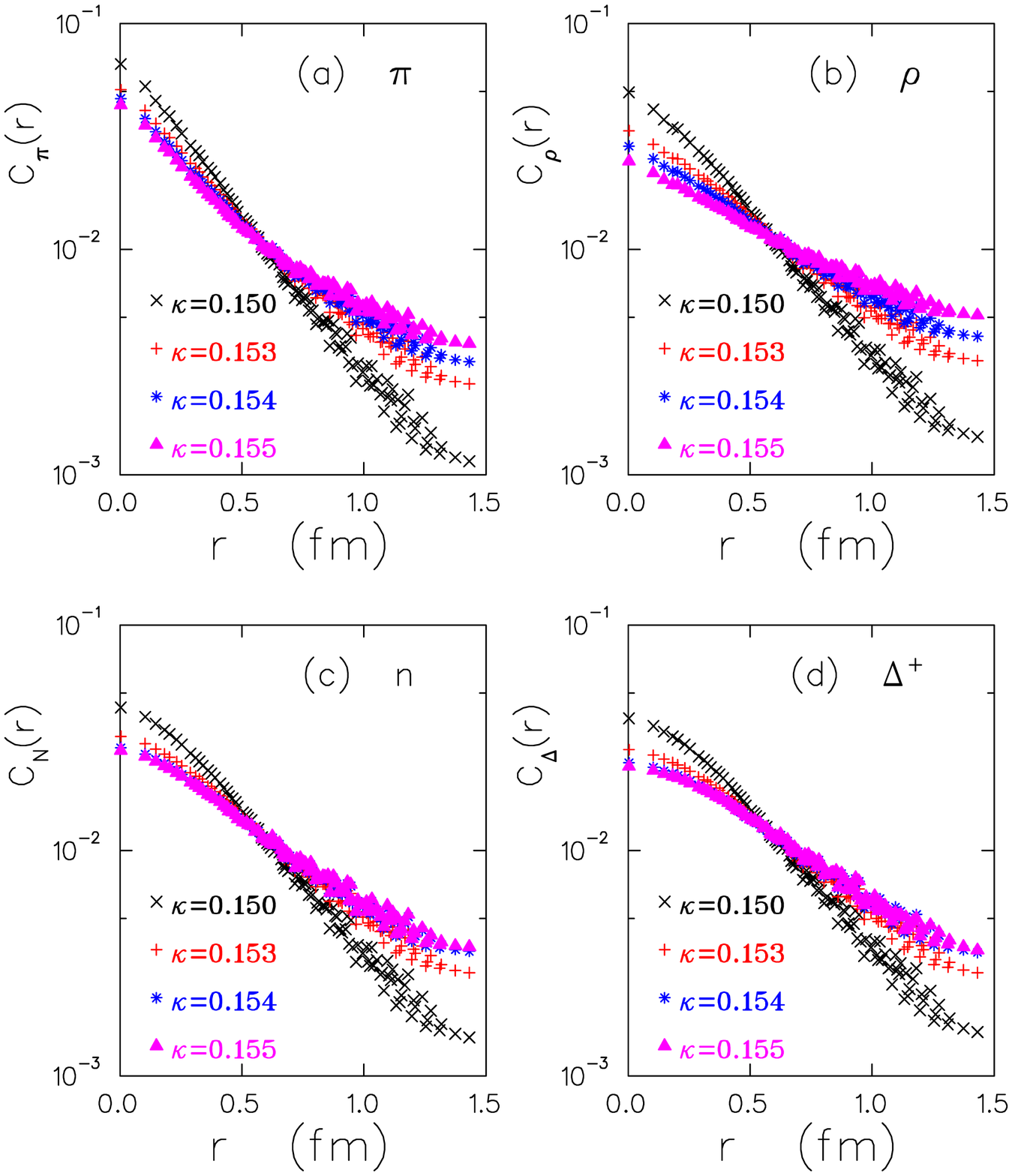}}
\vspace*{-1.5cm}
\caption{Density-density correlators as a function of the quark mass.} 
\label{fig:wfs}
\end{minipage}
\hspace{\fill}
\begin{minipage}[b]{7.8cm}
\epsfxsize=8truecm
\epsfysize=7truecm
\mbox{\epsfbox{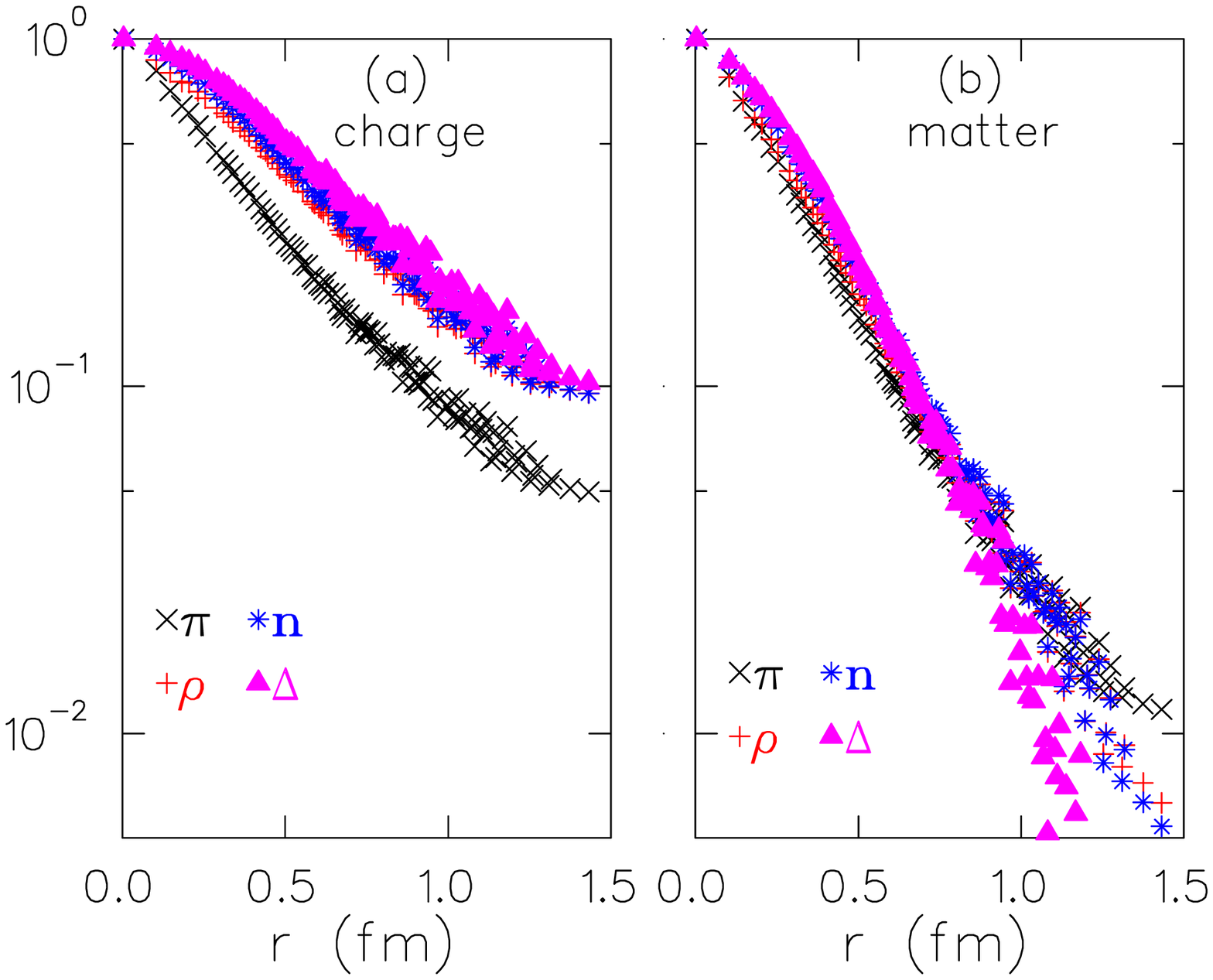}}
\vspace*{-1.5cm}
\caption{Comparison of charge (a) and matter (b) densities at $\kappa=0.153$.}
\label{fig:matter}
\end{minipage}
\end{figure}

The dependence of hadron wave functions on the
quark mass is displayed in Fig.~\ref{fig:wfs} where
 we plot the density-density correlators at various $\kappa$ values for the
pion, the rho, the nucleon and the $\Delta^+$ normalized to unity.
The size of the nucleon and $\Delta$ wave functions  do not change
 for naive quark mass smaller than $130$ MeV.
The rho wave function shows the largest variation with the  quark mass.
In Fig.\ref{fig:matter} we compare the matter and charge densities
at $\kappa=0.153$. We find that the matter density falls off
more rapidly than the charge density and it has the same shape
for all four hadrons.

\begin{figure}
\vspace*{-0.9cm}
\begin{minipage}[b]{7.cm}
\epsfxsize=7.truecm
\epsfysize=6truecm
\mbox{\epsfbox{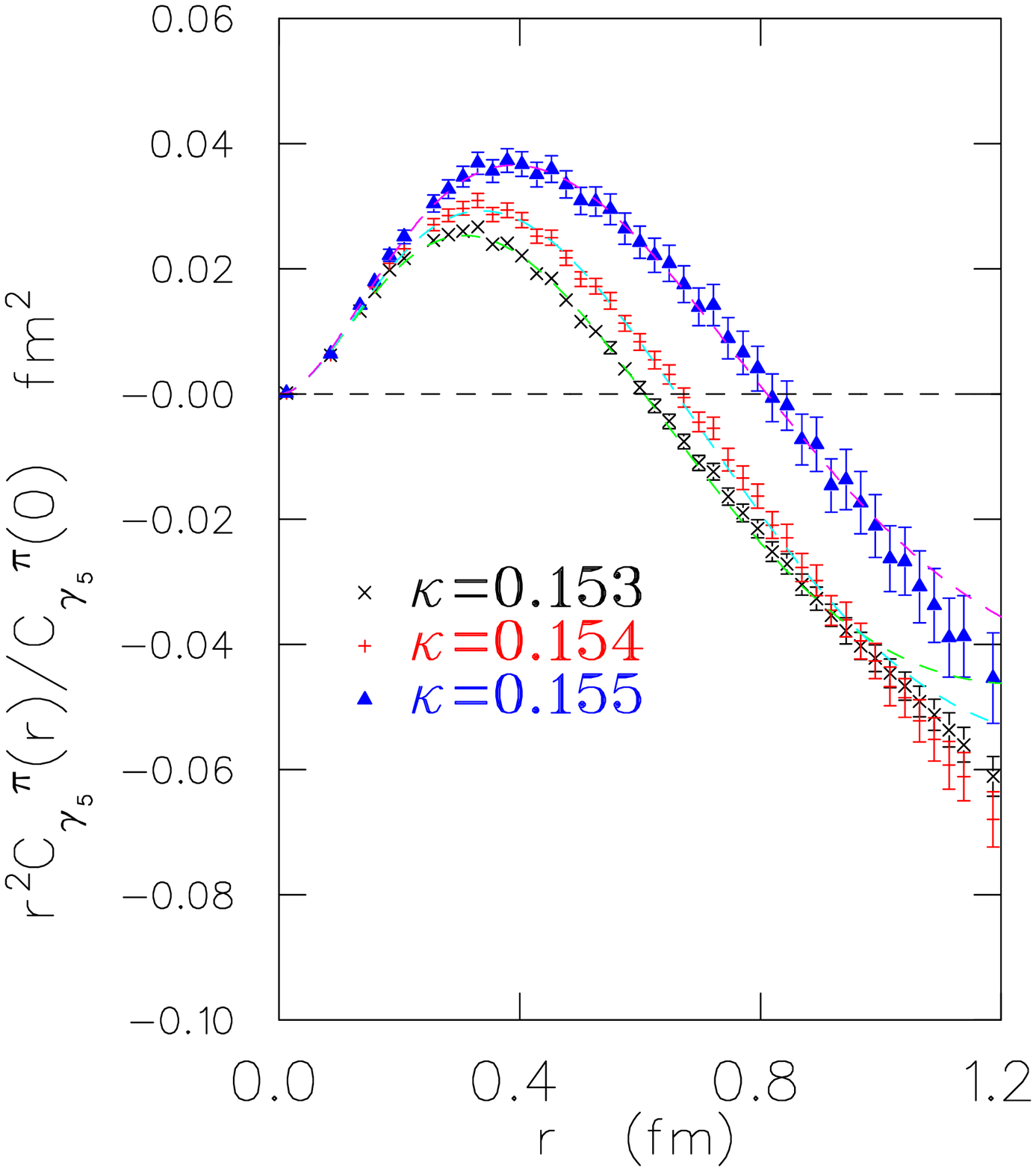}}
\vspace*{-1.3cm}
\caption{The pseudoscalar density for the pion 
at $\kappa=0.153, 0.154$ and 0.155 
with fits to $r^2(a+br^2+cr^4)\exp(-mr)$.}
\label{fig:pion55}
\end{minipage}
\hspace{\fill}
\begin{minipage}[b]{8.cm}
\epsfxsize=8truecm
\epsfysize=6truecm
\mbox{\epsfbox{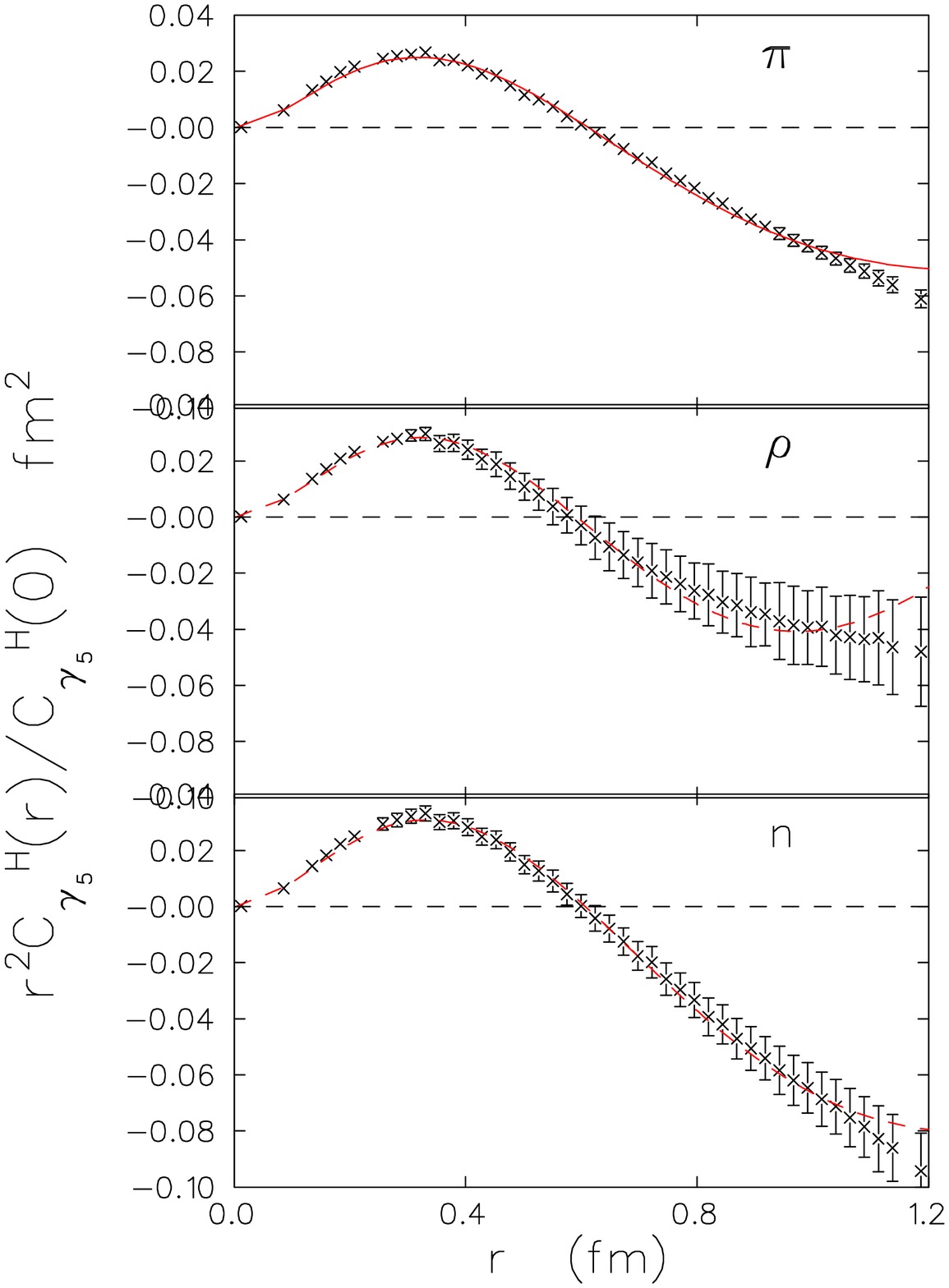}}
\vspace*{-1.3cm}
\caption{Pseudoscalar density  for the pion (upper), the rho (middle) and
the nucleon (lower) at $\kappa=0.153$. Fits as in Fig.~\ref{fig:pion55}.}
\label{fig:all55}
\end{minipage}
\vspace{-0.7cm}
\end{figure}

The pseudoscalar density for the pion is shown in Fig.~\ref{fig:pion55}
for $m_\pi/m_\rho= 0.84,\>0.78$ and $0.70$. In Fig.~\ref{fig:all55} it is
compared with that for the rho and the nucleon at $m_\pi/m_\rho=0.84$.
The bag model predicts that the integral $\int d^3r C_{\gamma_5}^H(r)$ is
 zero~\cite{Negele}. 
A reasonable fit is obtained by an exponential times a polynomial ansatz.
The long tail of the data and the integral of the fitted ansatz both
favor a non-zero integral. However a careful extrapolation to the continuum
limit (using large volumes) is required for this quantity, especially with Wilson fermions as used here.

The rho deformation  seen in Fig.~\ref{fig:rho it}
can be made more quantitative by analysing the density-density
correlator  into a dominant $L=0$ state
and a suppressed $L=2$~\cite{Gupta}:

\vspace*{-0.3cm}

\be <\rho_j({\bf 0})|\hat{\rho}_{\gamma_0}({\bf r})\hat{\rho}_{\gamma_0}({\bf 0})|\rho_j({\bf 0})> 
= \phi_1(r) + \frac{(3x_j^2-r^2)}{3r^2} \phi_2(r)
\label{angular}
\ee
\vspace*{-0.3cm}
where $|\rho_j({\bf 0})>$ is a zero momentum state with polarization $j$.

\begin{figure}
\vspace*{0cm}
\begin{minipage}{8.3cm}
\epsfxsize=9.truecm
\epsfysize=6.5truecm
\mbox{\epsfbox{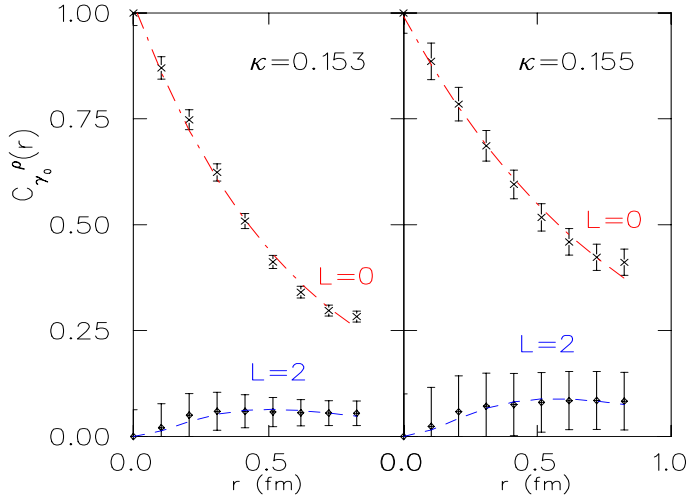}}
\vspace*{-1cm}
\caption{Decomposition of $C_{\gamma_0}$ for the rho into angular momentum part $L=0$ and $L=2$.}
\label{fig:angular}
\end{minipage}
\hspace{\fill}
\begin{minipage}{6.5cm}
As shown in Fig.~\ref{fig:angular}, a
 good description of the rho correlator,
is obtained by taking $\phi_1(r) = A \exp(-m_1 r) $ and 
$\phi_2(r) = B r^2 \exp(-m_2 r)$. 
 Using the values extracted from the fits and neglecting $B^2$ terms we find
for the deformation
$\delta\equiv (3/4)(\langle 3z^2-r^2 \rangle/\langle r^2 \rangle)\sim 0.01$ with an error of about 80\% which mainly arises from the 
poor determination of the coefficient of the $L=2$ state.
A direct determination of the quadrupole
moment from the rho correlator yields $\delta=0.03 \pm 0.01$ in reasonable agreement
with the value obtained from the angular decomposition.
\end{minipage}
\vspace*{-1cm}
\end{figure}

First lattice results for the ratio  of the  electric quadrupole to magnetic dipole amplitudes, $R_{EM}$, for both quenched and for two dynamical
Wilson fermions are  obtained at the smallest allowed lattice 
 momentum,  ${\bf q}=(2\pi/L_s,0,0)$, where $L_s$ is the spatial length of the lattice.
 At ${\bf q}^2= 0.54$~GeV$^2$  for the unquenched theory 
using the SESAM configurations~\cite{SESAM} we find in the chiral limit
$R_{EM}\sim (-3.7 \pm 0.5)\%$. 
For ${\bf q}^2 = 0.15$~GeV$^2$ in the quenched theory on  a lattice
of size  $32^3\times 64$  at $\beta=6.0$ we find  $R_{EM} \sim (-3.0\pm 0.3)\%$.
We used the nucleon mass to set the physical scale. 
Finite
lattice spacing effects on the $R_{EM}$ ratio can be significant
and are currently under investigation.  The finite volume dependence
on $R_{EM}$ is also under study.

\vspace*{-0.5cm}

\section{Conclusions}

\vspace*{-0.3cm}

We have presented a gauge-invariant determination of hadron profiles in the
quenched approximation.
We have found that  the rho wave function 
shows the strongest dependence on the quark mass whereas the nucleon and $\Delta$ the weakest.
We have established that the rho is deformed  
 with deformation 
which increases as we approach
the chiral limit, 
whereas the $\Delta^+$ has no statistically significant 
deformation. The matter density is similar for all four hadrons and
falls off more rapidly than the charge density.
Using state-of-the-art lattice techniques 
the phenomenologically 
important {$R_{EM}$}  has been  extracted for two  $q^2$ values and
found to be
consistent with 
experimental measurements.

\end{document}